\documentclass[sigconf, 10pt]{acmart}

\AtBeginDocument{%
  \providecommand\BibTeX{{%
    \normalfont B\kern-0.5em{\scshape i\kern-0.25em b}\kern-0.8em\TeX}}}

\setcopyright{acmcopyright}
\copyrightyear{2022}
\acmYear{2022}
\acmDOI{XXXXXXX.XXXXXXX}

 \acmConference[MobiArch]{Workshop on Mobility in the Evolving Internet Architecture}{October 21,
   2022}{Sydney, Australia}
\usepackage{amsmath}
\usepackage{multirow}
\usepackage{array}
\usepackage{pifont}
\newcommand{\cmark}{\ding{51}}%
\newcommand{\xmark}{\ding{55}}%

\newcolumntype{C}[1]{>{\centering\let\newline\\\arraybackslash\hspace{0pt}}m{#1}}

\begin{document}

\title{A Survey on Participant Selection for Federated Learning in Mobile Networks}

\newcommand{\tsc}[1]{\textsuperscript{#1}} 
\author{Behnaz Soltani\tsc{1}, Venus Haghighi\tsc{1}, Adnan Mahmood\tsc{1}, Quan Z. Sheng\tsc{1},  Lina Yao\tsc{2}}
\affiliation{
  \institution{\tsc{1}Macquarie University, Sydney, Australia}
  \institution{\tsc{2}University of New South Wales, Sydney, Australia}
  \country{}
}

\begin{abstract}
 Federated Learning (FL) is an efficient distributed machine learning paradigm that employs private datasets in a privacy-preserving manner. The main challenges of FL are that end devices usually possess various computation and communication capabilities and their training data are not independent and identically distributed (non-IID). Due to limited communication bandwidth and unstable availability of such devices in a mobile network, only a fraction of end devices (also referred to as the \emph{participants} or \emph{clients} in a FL process) can be selected in each round. 
 Hence, it is of paramount importance to utilize an efficient participant selection scheme to maximize the performance of FL including final model accuracy and training time. In this paper, we provide a review of participant selection techniques for FL. First, we introduce FL and highlight the main challenges during participant selection. Then, we 
 review 
 the 
 existing studies and categorize them based on their solutions. Finally, we provide some future directions on participant selection for FL based on our analysis of the state-of-the-art in this topic area.
\end{abstract}

\ccsdesc[500]{Computing methodologies~Machine learning}
\ccsdesc[500]{Computing methodologies~Distributed computing methodologies}

\keywords{Federated learning, machine learning, participant selection.}

\maketitle

\section{Introduction}

Mobile devices such as vehicles and smart phones are constantly generating a massive amount of data, which could be utilized for machine learning in a bid to achieve smart mobile applications. However, transmitting private data to a centralized or an edge server for training may lead to privacy issue and could cause long communication latency and large resource cost~\cite{verbraeken2020survey}. 

A decentralized machine learning approach called Federated Learning (FL) has been proposed by Google that enables cooperative learning on devices without sharing the local data~\cite{mcmahan2017communication}. Clients train the model on-device in a privacy-preserving manner using their local datasets and transfer the local model parameters to the FL server for aggregation. 
As a result, FL enables user privacy preservation, low communication costs, and transmission latency reduction owing to transmitting only model parameters to the server for aggregation. Furthermore, in time-critical systems such as autonomous vehicles, making real time decisions locally at end devices significantly decreases response time. 

Employing various datasets of different clients using FL leads to model accuracy improvement. 
Participant selection is an emerging challenge in the management of FL that possesses a profound impact on the performance of the model training, especially in the scenarios with a huge number of participants and limited wireless channels. In FL, due to dynamic environments and the limited network bandwidth, only a fraction of clients can be participated for training in each round. Hence, proper selection is of paramount importance in FL to achieve the desired accuracy with fast convergence. 
In FedAvg~\cite{mcmahan2017communication} algorithm introduced by Google, participants are determined uniformly at random in each round. However, due to the various computation and communication resources and heterogeneous datasets, devices may have different contributions on the training performance. 
To evaluate the performance of FL, {\em training time} and {\em final model accuracy} are the most important factors. Mostly, there are trade-offs between the model accuracy and convergence time in FL. For example, participants with high quality data may have poor network connection or computation capacities, while those with low quality data may possess rich resources to perform training process.


To the best of our knowledge, there are no existing works that conduct a literature review of the participant selection in FL. The main contributions of this study are as follows:
\begin{itemize}
    \item We discuss the important challenges pertinent to participant selection process in FL.
    \item We review the state-of-the-art on participant selection by categorizing them vis-\`a-vis different approaches.
    \item We identify some open research directions of participant selection in mobile networks.
\end{itemize}

\section{Federated Learning}



Federation learning (FL), in essence, encompasses a number of steps which are delineated as follows:

\begin{enumerate}
\item \emph{Initialization} -- The aggregation server determines the FL task, generates a randomly or pretrained global model, and adjusts learning parameters (e.g., the number of rounds and the learning rates).
\item \emph{Participant Selection} -- The FL server selects $n$ number of clients amongst the volunteers to participate in the training process.
\item \emph{Global Distribution} -- The server disseminates the global model parameters to the selected participants to train and update the shared model.
\item \emph{Local Training} -- Selected participants train the shared model using their local data samples and upload local model parameters to the server.
\item \emph{Aggregation} -- After receiving the local models, the server aggregates them and computes a new global model.
\end{enumerate}

\begin{figure}[t]
\centering
\includegraphics[width=\linewidth]{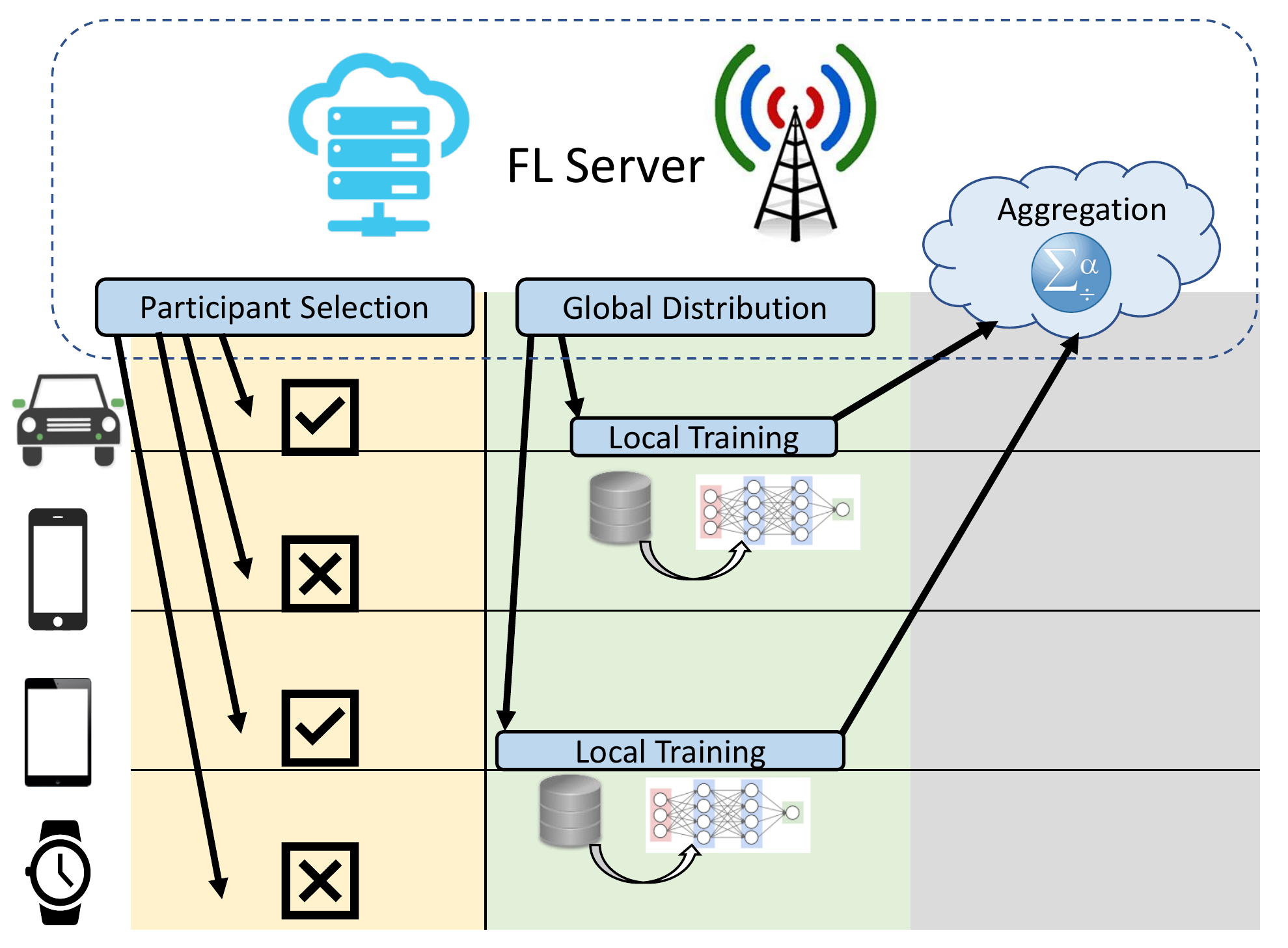}
\caption{An Overview of Federated Learning Process.}
\label{fig:FL}
\end{figure}

Steps 2, 3, and 4 repeat until the model converges or reaches a desirable accuracy and are portrayed in Figure~\ref{fig:FL}. The most popular FL algorithm is FedAvg~\cite{mcmahan2017communication} that randomly samples a subset of clients. However, due to different computation and communication capabilities and various data samples, randomly selected participants could degrade the performance of FL process.

\subsection{Challenges of Participant Selection}

There are several significant challenges in participant selection approaches that affect the performance of FL process:

\begin{itemize}
\item \textbf{Device Heterogeneity:} Devices have different computation, communication, and storage capabilities that may negatively affect the performance of training process~\cite{luo2021tackling}. In synchronous FL with heterogeneous resources, per-round training time is determined by the slowest clients (i.e., \emph{stragglers}) since all the participants are required to wait for the slowest clients to complete their training tasks.
On the other hand, end devices with limited resources may increase dropouts during training process and affect the accuracy of the model. Therefore, selecting clients without considering their resource capabilities may lead to longer training time and degrade the final model accuracy. 
\item \textbf{Data Heterogeneity}: Since data is generated independently based on clients' behaviours, data distribution on end devices are usually non-Independent and Identically Distributed (non-IID). Hence, local datasets may not be representative of the population distribution, which can lead to the biased model update and degrade the model accuracy~\cite{mcmahan2017communication,li2019convergence, wang2020optimizing, zhang2021client}. 
\item \textbf{Dynamicity:} Clients might be unavailable due to the high-mobility environment, poor network condition, and energy constraints~\cite{zhou2022role,kang2019incentive,kang2020reliable}. Furthermore, due to the channel fading in wireless networks, a fraction of local model updates may be lost. Therefore, a dynamic environment with high mobility devices has a major impact on the performance of FL process.
\item \textbf{Trustworthiness:} Since the FL server has no knowledge about the local training process, malicious devices may launch attacks and manipulate the result of the training task. Therefore, it is of utmost importance to identify malicious clients and eliminate them from the learning process.  
\item \textbf{Fairness:} Devices with poor capabilities are less likely to be selected to participate in the training process, which leads to selection bias and degrade the model accuracy~\cite{huang2022stochastic}. Fairness enables clients with diverse datasets to participate in the FL process and improve the model accuracy and convergence speed. Therefore, in order to minimize the model bias and generalize the global model, all the end devices should have a chance to participate in the FL process. 
\end{itemize}
%

\section{Participant selection in FL}

Over the past few years, researchers in both academia and industry have proposed a number of research studies pertinent to participant selection in the FL process. The same have been classified into eight
appropriate categories and are discussed as follows:

\vspace{1mm}
\noindent\textbf{Threshold-based Selection.} 
Participants may be allowed to complete the local training within a specific deadline.
FedCS~\cite{nishio2019client} is one of the first studies on client selection in FL that manages participant selection process based on the resource capabilities. The proposed method aims to select as many clients as possible that can complete the training steps within a specific deadline to reach the desired performance~\cite{mcmahan2017communication}. First, the random clients are asked to inform 
their resource information, i.e., computation and computation capacities, wireless channel states, and the size of their data. Afterward, the server computes the required time to complete the training task for each client using aforementioned information. 
 However, due to dynamically changing resource and network conditions, the static time threshold 
cannot guarantee the efficiency of this approach. Thus, an adaptive deadline determination algorithm for mobile devices is proposed in~\cite{lee2021adaptive}, 
wherein clients are adaptively determined at each round instead of using a fixed deadline. This 
approach could decrease the convergence time by up to 50\% 
in contrast to the conventional fixed threshold schemes. 

A multicriteria-based client selection strategy is proposed in~\cite{abdulrahman2020fedmccs} in which time, CPU, memory, and energy for all the end devices are considered to predict their 
respective capability to perform 
a training task. The authors employ stratified-based sampling to establish a homogeneous group of clients based on the time zones. Subsequently, 
considering the resource utilization and the resource capabilities to complete the FL task within a specific time frame, the authors 
formulate a bilevel optimization problem to maximize the number of selected clients. To predict the resource utilization, linear regression according to the history of the past rounds is employed. Furthermore, due to imbalanced class distribution, the authors prioritize participants with the highest event rate (i.e., the ratio of abnormal data samples pertaining to minority class to the total data samples) 

\vspace{1mm}
\noindent
\textbf{Reputation-based Selection.} Unreliable clients may (a) perform undesirable behaviours intentionally, 
e.g., data poisoning, model poisoning, and sybil attacks, 
or (b) unintentionally due to high mobility environments and unstable network connections. Therefore, it is of paramount importance to design an efficient participant selection approach for reliable model training. Reputation is a metric to evaluate trustworthiness of entities based on their past interactions~\cite{mahmood2021trust, mahmood2022trust, liu2011novel}. Recently, a couple of studies employed reputation as a metric to select reliable devices 
for participating in FL process. A reputation-based client selection scheme is proposed in~\cite{kang2019incentive,kang2020reliable} that employs multiweight subjective logic 
by taking into account both the direct interactions and recommendations 
from other task publishers. The proposed approach considers three weight attributes: 
interaction frequency, interaction timeliness, and interaction effects. The reputation 
is stored in a 
blockchain to achieve secure reputation management.

In~\cite{zou2021reputation}, a reputation-based regional FL framework is introduced 
for intelligent transportation systems in which vehicles are divided into multiple regions 
each of which possesses its own learning model. Roadside Units (RSUs) are selected as the leaders and perform aggregation in each region. 
Owing to the dynamic nature of transportation systems, devices may move away from their regions 
and join a new region. 
Each leader computes the reputation of vehicles based on a) honesty degree, b) accuracy contribution, and c) interaction timeliness, and selects vehicles with high reputation to join FL process. 
In~\cite{song2021reputation}, the authors introduce a reputation model based on beta distribution in order to measure the trustworthiness of the end devices. The trust value is evaluated using the contribution of participants to the global model. They propose a reputation-based scheduling scheme that jointly considers trustworthiness and fairness based on the reputation value and successful transmission rate.

\vspace{1mm}
\noindent\textbf{Probability Allocation-based Selection.}
Different end devices can have various probabilities of being selected for a training task. The authors in~\cite{luo2021tackling} design a client sampling strategy to reduce total training time. They take both system and data heterogeneity into consideration 
since clients with high-quality data may have poor communication resources, whereas, those with high communication capabilities may have low-quality data. Their approach provides a new convergence upper bound for arbitrary client selection probabilities and generates 
a non-convex training time minimization problem.
Their approach significantly reduces the convergence time for achieving the same target loss compared to several baselines.

In~\cite{huang2022stochastic}, a stochastic client selection algorithm under a volatile context is investigated in which the selected clients might not be capable of returning their local models for aggregation due to different reasons, 
including but not limited to, limited computing resources, network failure, and user unwillingness. However, selecting clients with the lowest failure probability may violate selection fairness owing to the selection of a particular group of clients repeatedly and hurt the model accuracy. Therefore, the authors study the trade-off between selecting participants with the lowest failure likelihood and selection fairness, and formulate the client selection problem by taking both factors into account. Since the problem cannot be solved offline due to unknown status of the participants, an adversary bandit-based online scheduling solution is employed. The proposed method is able to increase convergence time while maintaining the model accuracy level.

The authors in~\cite{perazzone2022communication} derive a novel convergence upper bound for non-convex loss functions using FL with arbitrary device selection probabilities. They design a stochastic optimization problem that aims to minimize a weighted sum of the convergence bound and communication time. The proposed participant selection method solves the problem using the Lyapunov drift-plus-penalty framework based on current channel conditions without the knowledge of channel statistics.
In~\cite{wu2022node}, an aggregation algorithm is designed to determine the optimal subset of local model updates by excluding adverse local updates. Moreover, a Probabilistic Node Selection framework (FedPNS) is proposed that dynamically adjusts the selection probability for devices based on their contribution to the model pertaining to the data distribution that is determined using the output of the aggregation algorithm.

\vspace{1mm}
\noindent\textbf{Reinforcement Learning-based Selection}. In reinforcement learning, an agent learns to attain an objective in a complex and uncertain environment and maximizes its rewards. A framework based on multi-armed bandit for online client selection is introduced in~\cite{xia2020multi} to minimize the training latency including both local computation time and data transmission time in two scenarios: 1) an ideal scenario in which clients possess balanced and IID datasets and are always available, and 2) a non-ideal scenario in which clients may be 
unavailable and the distribution of datasets is non-IID. In the non-IID scenario, both the fairness and availability constraint is addressed.

In~\cite{huang2020efficiency}, the authors define an offline client selection problem with long-term fairness constraints. The constraint adjusts the minimum average selection rate of every participant to maintain fairness for the system. Due to the indeterminate availability of clients until the start of a round and also time-coupling fairness constraint, the authors utilize the Lyapunov optimization framework to transfer the offline problem into an online optimization problem, where the participation rate of clients is evaluated using dynamic queues. In the proposed method, the model exchange time (i.e., the time spent between global model distribution and uploading all the local models) of each participant before each communication round is estimated using Contextual Combinatorial Multi Arm Bandit (C$^2$MAB) model. The authors aim to minimize the long-term model exchange time.

An experience-driven controlled framework called FAVOR is proposed in~\cite{wang2020optimizing} that aims to determine the best fraction of devices in each round to minimize the number of rounds and tackle the data non-IID distribution. The authors formulate client selection for FL process as a deep reinforcement learning problem. To select devices in each communication round, a double deep Q-learning mechanism is proposed in order to improve global model accuracy and reduce the number of communication rounds. 

\vspace{1mm}
\noindent\textbf{Group-based Selection (GS).} Clients are divided into several groups and then those within the same group 
are 
selected for each communication round. In~\cite{ma2021client}, a grouping based participant selection mechanism is introduced in which participants are split into various groups based on group earth mover’s distance (GEMD) to balance the label distribution of the clients. This new metric evaluates similarity between global distribution and local data distributions. A smaller GEMD means that the training data of the selected clients are closer to IID distribution. Therefore, selecting a group of clients with the smallest GEMD can improve the performance of FL.

\begin{table*}[h!]

\centering
 \caption{Comparison of the existing participant selection works.}
 \resizebox{\linewidth}{!}{
 \begin{tabular}{||c||C{3cm}|c|c|c|c|c|c||} 
 \hline
 No. & Methods  & Device Heterogeneity & Data Heterogeneity & Fairness & Dynamicity & Trustworthiness& Objective\\ 
 \hline\hline
\cite{nishio2019client} & 
\multirow{2}{3cm}{\centering Threshold-based}&\cmark & \xmark & \xmark & \xmark & \xmark & Maximizing the number of participants\\ \cline{1-1}\cline{3-8}
\cite{lee2021adaptive} & &\cmark & \xmark & \xmark & \cmark&\xmark & Maximizing the number of participants\\\cline{1-1}\cline{3-8} 
\cite{abdulrahman2020fedmccs} & &\cmark & \cmark & \xmark & \xmark & \xmark& Maximizing the number of participants\\\hline 

\cite{kang2019incentive,kang2020reliable} &
 \multirow{4}{3cm}{\centering Reputation-based}&\xmark & \xmark & \xmark & \xmark & \cmark& Slecting trusted participants to improve the reliablity of FL \\ \cline{1-1}\cline{3-8} 
\cite{zou2021reputation} & &\xmark & \xmark & \xmark & \cmark & \cmark & Reliable participant selection to improve accuracy of knowledge\\ \cline{1-1}\cline{3-8} 

\cite{song2021reputation} & &\xmark & \xmark & \cmark & \cmark & \cmark& Improving the reliability and convergence performance \\\hline
 
\cite{luo2021tackling} &
 \multirow{4}{3cm}{\centering Probability allocation-based}&\cmark & \cmark &  \xmark &  \xmark &  \xmark & Minimizing total learning time \\\cline{1-1}\cline{3-8} 
\cite{huang2022stochastic} & &\cmark & \xmark &  \cmark &  \cmark &  \xmark & Addressing the trade-off between the lowest failure
probability and fairness  \\\cline{1-1}\cline{3-8} 
\cite{perazzone2022communication}& &\cmark & \cmark &  \xmark &  \cmark &  \xmark & Minimizing communication time for speeding up the convergence\\\cline{1-1}\cline{3-8} 
\cite{wu2022node} & &\xmark & \cmark & \xmark &  \xmark &  \xmark & Excluding adverse local models from participating devices\\\hline

\cite{xia2020multi} &
 \multirow{3}{3cm}{\centering Reinforcement learning-based}&\cmark  & \cmark & \cmark & \cmark & \xmark & Minimizing the total training time \\\cline{1-1}\cline{3-8} 
\cite{huang2020efficiency} &&\cmark& \xmark  &\cmark &\cmark & \xmark & Minimizing exchange time with fairness guarantee \\\cline{1-1}\cline{3-8} 
\cite{wang2020optimizing} & &\xmark & \cmark & \xmark & \xmark& \xmark & Reducing the number of communication round under non-IID setting\\\hline
  
\cite{ma2021client} &Group-based&\xmark & \cmark  &\xmark &\xmark & \xmark & Balancing the label distribution of participants\\\hline

\cite{zhao2022participant} &
 \multirow{2}{3cm}{\centering Weight Divergence-based}&\xmark & \cmark  &\xmark &\xmark & \xmark& Addressing trade-off between accuracy contribution and training time \\\cline{1-1}\cline{3-8} 
\cite{zhang2021client}& &\xmark & \cmark  &\xmark &\xmark & \xmark & Selecting clients with lower non-IID degree of data \\\hline
 
\cite{wang2021device}&Offloading-Aware&  \cmark & \cmark & \xmark & \xmark & \xmark& The optimal combination of client selection and data offloading configuration\\\hline

\cite{fraboni2021clustered} & Clustering-based &\xmark & \cmark &  \xmark &  \xmark &  \xmark & Unbiased participant selection \\
 \hline

 \end{tabular}
} 
 \label{table:1}
\end{table*}

\vspace{1mm}
\noindent\textbf{Weight Divergence-based Selection.} Weight divergence in training can be an indicator to identify the distribution of local datasets. The authors in \cite{zhao2022participant} introduce a client selection utility that tries to deal with the trade-off between accuracy and execution time in each round. The change of weight between two adjacent rounds 
is 
defined as a utility for fast convergence. In addition, since clients with large data volume may negatively affect the training time, 
the ratio of the local data size to the total size
is also added as a coefficient to the client's utility in order that if local data constitutes a significant amount of total data, utility reduces. 
Since it is not always necessary to select participants in every round, this study also designs a feedback control component that dynamically adjusts the frequency of client selection.

The authors in~\cite{zhang2021client} design a participant selection algorithm to tackle the accuracy reduction owing to non-IID datasets. They identify the degree of non-IID data using the weight divergence. The weight divergence is evaluated between the client model and auxiliary model 
which is trained using public or purchased datasets at the aggregation server. 
In this approach, participants that have lower non-IID degree should be selected with higher frequency for training.

\vspace{1mm}
\noindent\textbf{Offloading-Aware Selection.} Device-to-device (D2D) offloading of local data processing from resource-limited devices to resource-rich devices can enable local training with more diversified data distribution. Participant sampling with data offloading are combined in~\cite{wang2021device} to maximize training accuracy. 
Devices with large contributions to model training are chosen for training and other devices may send their data to the selected participants based on their data similarity. Two data samples are considered similar if they have identical labels and their feature vectors have limited difference. Data offloading is only performed between trusted and single-hop neighbors. To determine participants, the proposed method uses graph convolutional networks (GCNs) to learn the relationship between FL accuracy, network attributes, and offloading topology, which maximizes training accuracy.

\vspace{1mm}
\noindent\textbf{Clustering-based Selection.} Various clients can be selected with different data distributions using clustered sampling. In~\cite{fraboni2021clustered}, an unbiased clustered sampling scheme for client selection is introduced that decreases weights' variance for the client aggregation and ensures that clients with unique distributions are more likely to be selected. The authors introduce two clustered sampling schemes: 1) clustered sampling based on sample size, and 2) clustered sampling based on similarity. They experimentally show that clustered sampling leads to better and faster convergence.

In Table~\ref{table:1}, we summarize the existing studies and their approaches, and draw a comparison of them in terms of solving different challenges.  

\section{Future Research Directions}
Participant selection in FL is still in its infancy and there are 
some open challenges that require to be addressed. We have identified the following 
main research directions:

\vspace{1mm}
\noindent\textbf{Optimal and Adaptive Reputation Threshold.}  Existing reputation-based studies consider a pre-defined threshold in order to distinguish malicious devices from honest ones. 
If this threshold is adjusted too high, honest devices may lose the opportunity to participate in the FL process. On the other hand, if the aforementioned threshold is set too low, malicious devices may be selected for training and manipulate the model. It is, therefore, essential to optimize the threshold value, particularly, in safety-critical applications. 

\vspace{1mm}



\noindent\textbf{Hierarchical Aggregation.} All the reviewed approaches use centralized aggregation, wherein clients are required to transmit their local model parameters to a single aggregator server. When the number of devices increase, limited network bandwidth becomes a scalability bottleneck in centralized FL. On the other hand, communication between end devices and a remote server might be intermittent or unavailable.
 In hierarchical FL, end devices are divided into a number of clusters and participants transmit their local models to the cluster head for intermediate model aggregation. All cluster heads communicate with the FL server for global aggregation. This approach reduces the device dropout and improves scalability. To reduce communication cost and improve performance of FL, an efficient mechanism for determining the cluster structures is of paramount importance.


\vspace{1mm}
\noindent\textbf{Highly Dynamic Environment.} In mobile networks, i.e., the Internet of Vehicles with high mobility nodes and a highly dynamic topology, intermittent connectivity of devices with the FL server along with a constant change in data transmission time may degrade the performance of FL process. After client selection, some participants may get out of network coverage and significantly affect the system behaviour. It is important to handle high mobility environment and minimize the training performance degradation.
Decentralized FL can be a proper solution for high mobility clients that may 
not have access to the aggregation server. In decentralized FL, clients can only share their local model parameters with nearby devices using device-to-device communication without relying on a single server~\cite{barbieri2022decentralized}. Device selection algorithms in decentralized FL can have a significant impact on the performance of FL performance. Social trust values can be integrated with other selection criteria to determine the best nearby clients. 

\noindent\textbf{Asynchronous Training.} All existing works are based on synchronous training, wherein the aggregation server has to wait for receiving all the local models before aggregation. Therefore, \emph{stragglers} prolong the training time owing to data and device heterogeneity. In asynchronous FL, the server is not affected by \emph{stragglers} and can update the global model without waiting to collect all the local models. Participant selection in asynchronous FL needs to be studied so that appropriate selection approaches are proposed.


\section{Conclusion}
Participant selection in federated learning (FL) has a significant impact on the final model accuracy and training time of FL. 
In this paper, 
we explore the main FL challenges of the participant selection process. 
We review the existing approaches and conduct a comparison of the existing studies in terms of solving different challenges. Finally, we identify some research directions and hope to stimulate further research in this important topic. 


\bibliographystyle{unsrt}
\bibliography{sample-base}

\end{document}